\documentclass[superscriptaddress,twocolumn,prl]{revtex4-1}

\usepackage{graphicx}
\usepackage{amsmath}
\usepackage{amssymb}
\usepackage{amsthm}

\begin{document}

\title{A direct view at excess electrons in TiO$_2$ rutile and anatase}

\author{Martin Setvin}
\email{setvin@iap.tuwien.ac.at}
\affiliation{Institute of Applied Physics, 
Vienna University of Technology, Wiedner Hauptstrasse 8-10/134, 1040 Vienna, Austria}
\author{Cesare Franchini}
\email{cesare.franchini@univie.ac.at}
\affiliation{Faculty of Physics and Center 
for Computational Materials Science, Universit\"at Wien, Sensengasse 8/8-12, A-1090 
Wien, Austria}
\author{Xianfeng Hao}
\affiliation{Institute of Applied Physics, 
Vienna University of Technology, Wiedner Hauptstrasse 8-10/134, 1040 Vienna, Austria}
\author{Michael Schmid} 
\affiliation{Institute of Applied Physics, 
Vienna University of Technology, Wiedner Hauptstrasse 8-10/134, 1040 Vienna, Austria}
\author{Anderson Janotti}
\affiliation{Materials Department, University 
of California, Santa Barbara, CA 93106-5050}
\author{Merzuk Kaltak}
\affiliation{Faculty of Physics and Center 
for Computational Materials Science, Universit\"at Wien, Sensengasse 8/8-12, A-1090 
Wien, Austria}
\author{Chris G. Van de Walle}
\affiliation{Materials Department, University 
of California, Santa Barbara, CA 93106-5050}
\author{Georg Kresse}
\affiliation{Faculty of Physics and Center 
for Computational Materials Science, Universit\"at Wien, Sensengasse 8/8-12, A-1090 
Wien, Austria}
\author{Ulrike Diebold}
\affiliation{Institute of Applied Physics, 
Vienna University of Technology, Wiedner Hauptstrasse 8-10/134, 1040 Vienna, Austria}

\begin{abstract} 
A combination of Scanning Tunneling Microscopy/Spectroscopy 
and Density Functional Theory (DFT+\textit{U}) 
is used to characterize excess electrons in  TiO$_2$ rutile and anatase, 
two prototypical materials with identical chemical composition but different crystal lattices.
In rutile, excess electrons can localize at any lattice Ti atom, forming a small polaron, which can easily hop to neighboring sites.
In contrast, electrons in anatase prefer a free-carrier state, and can only be trapped 
near oxygen vacancies or form shallow donor states bound to Nb dopants.
The present study conclusively explains the differences between the two polymorphs and
indicates that even small structural variations in the crystal lattice can lead to a very different behavior.
\end{abstract}

\maketitle

The behavior of charge carriers in oxides is of key importance 
in virtually all applications of these materials. When excess electrons are added 
to the conduction band of an oxide, they may either retain the free-carrier character or, still assuming a defect-free crystal, couple to lattice distortions induced by its presence (electron-phonon interaction). The latter case is usually referred as to a small or large polaron, depending on the degree of electron localization~\cite{1,2}.
Polaronic effects and electron localization affect a materials' physical and chemical properties, 
yet it remains controversial how to model it appropriately from first principles \cite{3,4}.
Here we investigate TiO$_2$, a prototypical metal oxide.
TiO$_2$ is used in catalysis \cite{5,6,7,8}, photoelectrochemical
(Gr\"atzel) solar cells, memristors \cite{9}, and as a 
transparent conductive oxide \cite{10}. Two forms of TiO$_2$ are used industrially, rutile and anatase. 
The metastable anatase form is generally present in nanomaterials and shows better 
performance in energy-related applications and in optoelectronics. 
Even after several decades of research, a consensus on the origin
of the difference between the two materials is still absent, and our aim is to resolve this 
issue theoretically as well as experimentally.

Stoichiometric rutile and anatase are both insulators 
with a $\approx$3 eV band gap. TiO$_2$ can be turned into an \textit{n}-type 
semiconductor by adding excess electrons by various means - doping, UV irradiation, 
or chemical reduction. Electrons in the conduction band of TiO$_2$ 
compete between free-carrier and polaronic configurations. The extent 
to which this happens has remained highly controversial, yet strongly affects the 
material's transport properties and catalytic activity. The electrons can localize 
at Ti 3\textit{d} orbitals, forming Ti$^{3+}$ ions. This induces relaxations of the surrounding lattice atoms by typically 
0.1 \AA. The quasiparticle consisting of an electron coupled to the lattice relaxations 
in its immediate surrounding is called a small polaron \cite{1}. 
When thermally activated, small polarons exhibit hopping mobility. If the structural deformation spreads over a large number of lattice sites the corresponding solution is categorized as a large polaron.

Here we use the following joint theoretical and experimental approach: First, we theoretically investigate the intrinsic behavior of an excess electron added to the perfect crystal, i.~e. stoichiometric bulk cells of anatase and rutile. We establish that rutile allows polaron formation at any Ti site, while anatase prefers a free-carrier configuration. Next we inspect the effect of excess electrons donated by surface oxygen vacancies (V$_\mathrm{O}$s) by comparing experimental and DFT+U data. We use bulk-terminated rutile~(110)-(1$\times$1) and anatase~(101)-(1$\times$1) surfaces.  In rutile, the excess electrons leave the V$_\mathrm{O}$s and form polarons, which can hop through the lattice. In anatase, the electrons stay trapped at the V$_\mathrm{O}$s. Finally, we show that in Nb-doped anatase electrons are spatially confined by the donor potential, yet they keep the band-like character.

For our calculations we used the VASP code \cite{VASP}. On the experimental side, low-temperature Scanning Tunneling Microscopy/Spectroscopy (STM/STS) was used. Filled-states STM images reflect the spatial distribution of electrons within the bangap, and STS provides information about the electronic energy $E_\mathrm{EL}$ (see below): Either $\sim 1$~eV below $E_\mathrm{F}$ typical for small polarons \cite{5}, or $\sim40$~meV below $E_\mathrm{F}$ for delocalized, weakly bound electrons \cite{19}.

\begin{figure}[t!]
    \begin{center}
        \includegraphics[width=0.9\columnwidth,clip=true]{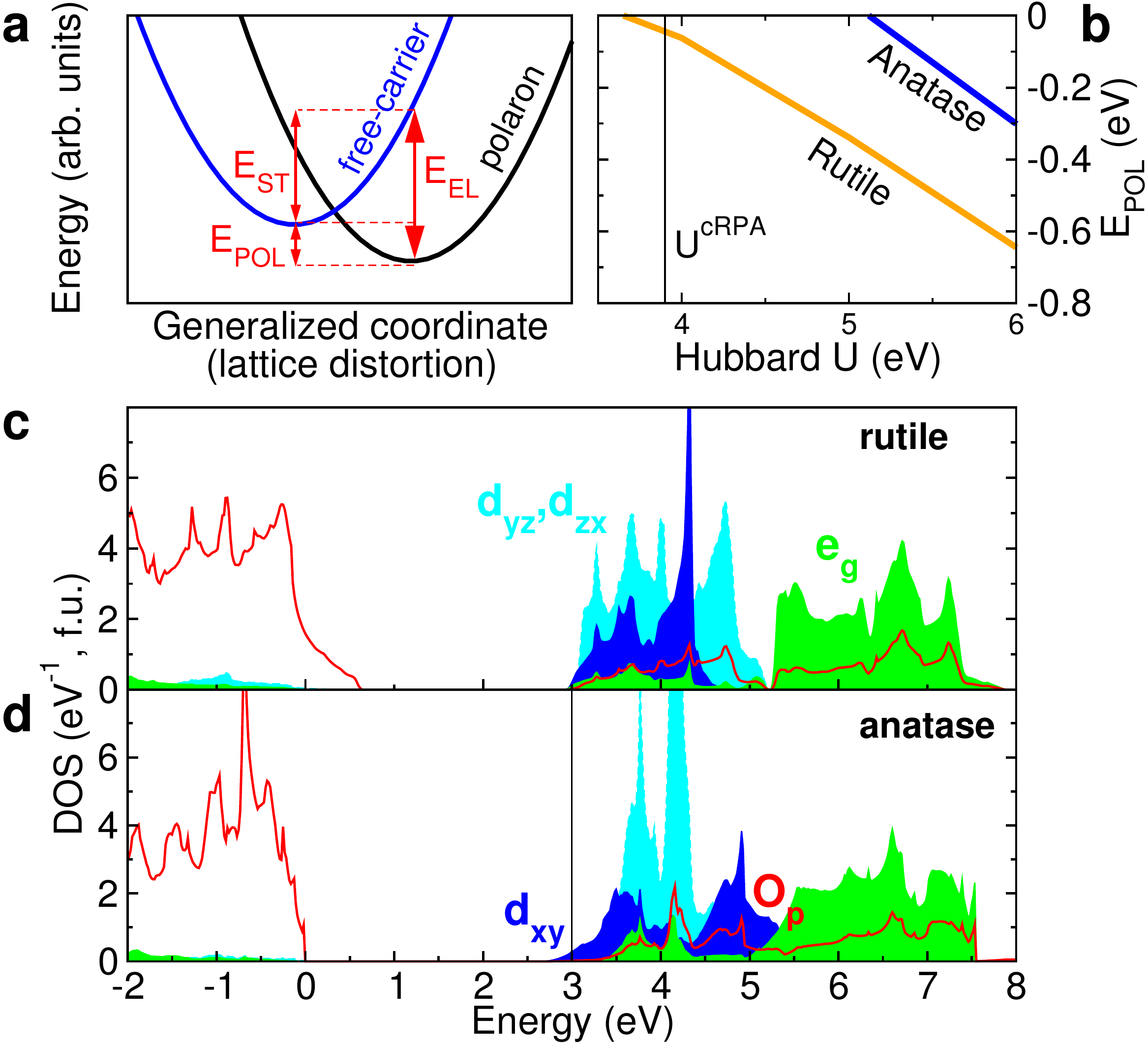}
    \end{center}
\caption{Calculated polaronic stability for bulk TiO$_2$ rutile and anatase.
a) Configuration coordinate diagram showing the polaronic ($E_{\rm POL}$),
lattice ($E_{\rm ST}$), and  electronic ($E_{\rm EL}$) energies as a function of lattice distortion
for the polaronic and delocalized solution.  b) $E_{\rm POL}$ as a function of Hubbard $U$ in bulk rutile (orange) and anatase (blue). The vertical line
indicates the {\em ab~initio} $U^\mathrm{cRPA}$. 
Orbitally decomposed density of states (DOS) in rutile (c) and anatase (d),
aligned with respect to the Ti core levels.}
\end{figure}
 
The energy balance for polaron formation in defect-free TiO$_2$ is sketched in Fig. 1(a). The formation 
energy, $E_{\rm POL}$, is 
defined as the total energy difference between the polaronic and fully delocalized
free-carrier solution, and results from the competition between the strain energy required 
to distort the lattice ($E_{\rm ST}$), 
and the electronic energy gained by localizing the electron at a Ti site in such a
distorted lattice, $E_{\rm EL}$. Polarons show a complex behavior. The electronic energy $E_{\rm EL}=E_{\rm POL}+E_{\rm ST}$
is the quantity measured in photoemission spectroscopy (PES) or STS, as the lattice atoms are ``frozen'' within the timescale
of the experimental probe \cite{2,5,11}. 
For the purpose of electrical conductivity, however, activation
energies are typically $\sim$tens of meV.  This is indicative of either
a low barrier for hopping between neighboring polaronic configurations, or a small
excitation energy from the polaronic to the free-carrier state \cite{11}.

Figure 1(b) illustrates why theoretical modeling of the polarons 
in TiO$_2$ and other oxides remains a challenging 
and controversial issue.  
Standard DFT always yields delocalized solutions.  Electrons 
can be localized by applying a Hubbard \textit{U}; the value 
of \textit{U}, and thus $E_{\rm EL}$, 
will then determine whether the polaronic solution will be stable \cite{12,13,14,15,16}. 
The situation is equally critical with hybrid functionals, where $E_{\rm EL}$ 
depends on the amount of exact exchange incorporated in the DFT functional \cite{4,11,17,18}.

From Fig. 1(b) we infer that $E_{\rm POL}$ 
is 0.4 eV larger in rutile than anatase, and that a larger \textit{U} 
is required to form a small polaron in anatase ($U>5$~eV) than in rutile ($U>3.5$~eV). 
We find that both materials have a similar $E_{\rm ST}$ (0.41~eV); the difference in $E_{\rm POL}$ 
originates from the electronic energy $E_{\rm EL}$. 
The formation of a polaron involves the depletion of the conduction band minimum 
(CBM), which has a different character in the two TiO$_2$ 
polymorphs. Figure~1~c,d shows that the conduction band in anatase is 1~eV wider 
than in rutile, and the CBM lies lower in energy as a result of the formation of 
a strong bonding linear combination between neighboring Ti $d_{xy}$ 
orbitals. Thus, a larger $U$ is required to alter this 
energetically favorable configuration and to form a polaron in anatase.  The 
associated energy gain $E_{\rm EL}$ is lower as compared to rutile. 

The most commonly used values of $U$ for TiO$_2$ 
range between 2.5 and 4.5~eV \cite{Cheng2009, Finazzi2008}.
We calculated the $U$ parameter entirely from first
principles using the constrained Random Phase Approximation (cRPA) \cite{cRPA}
and obtained a value of $U^\mathrm{cRPA}=3.9$~eV for rutile and 4.1 eV for anatase.
All results in the present work have been consistently determined using $U^\mathrm{cRPA}$=3.9 eV.
From Fig. 1, it is clear that $U^\mathrm{cRPA}=3.9$~eV suffices to stabilize the polaron
in rutile, albeit just, whereas polaron formation is clearly unfavorable in a perfect anatase lattice.
In rutile the excess electron is trapped in a Ti$^{3+}$ site, forming a small polaron;
the 6 O atoms surrounding Ti$^{3+}$ relax outward by 2-4 \%
of the equilibrium Ti--O bond length. In anatase the excess electron exhibits a free-carrier 'delocalized'
character: the crystal remains unperturbed and the excess electron is homogeneously distributed in 
the crystal.


\begin{figure}
    \begin{center}
        \includegraphics[width=1.0\columnwidth,clip=true]{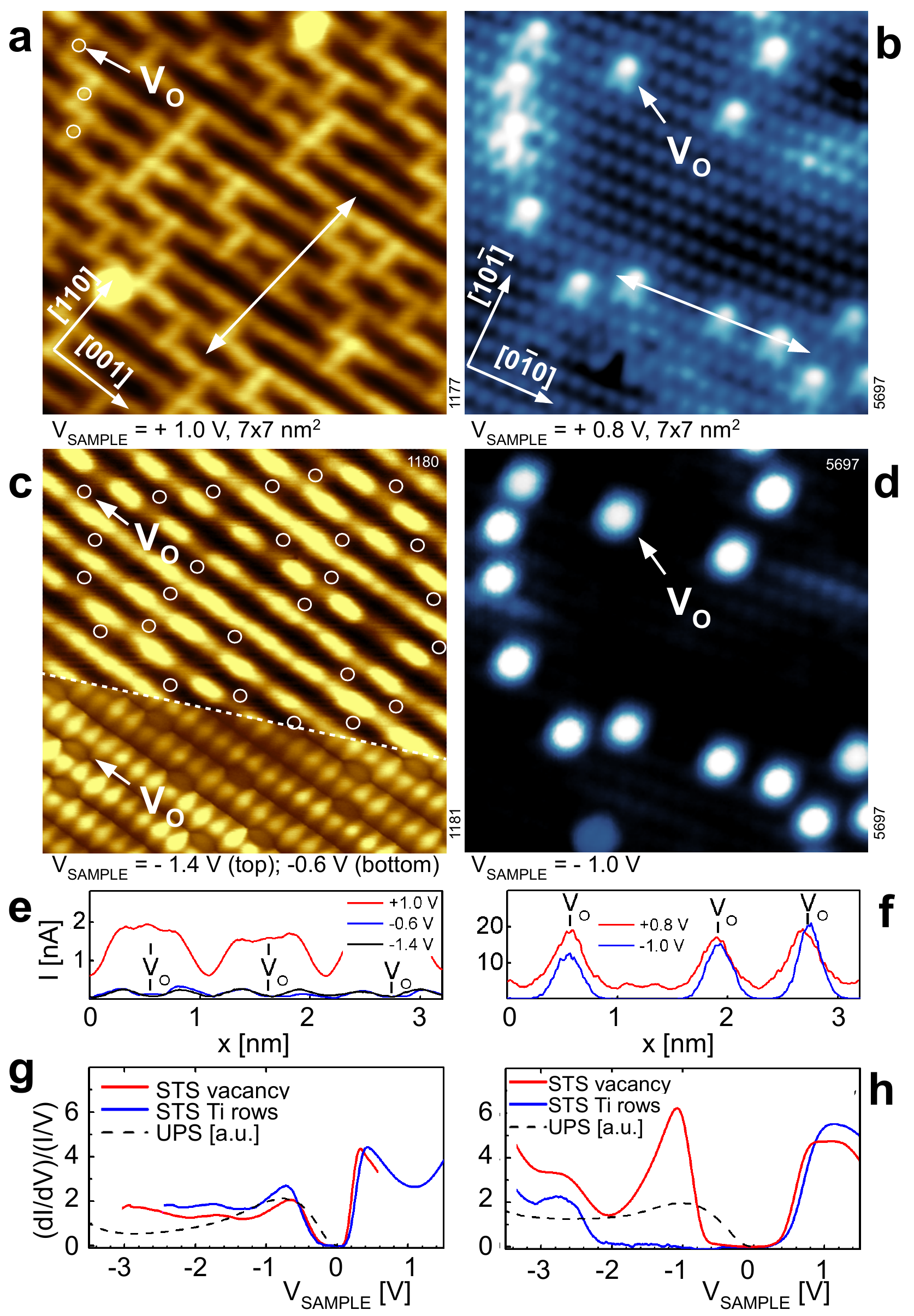}
    \end{center}
    \caption{
Local electronic structure of rutile (110) (left) and anatase (101)
(right) surfaces doped by surface oxygen vacancies. Constant-height scanning tunneling microscopy images of (a,
b) empty states and (c, d) filled states of the same areas.  V$_\mathrm{O}$ marks surface oxygen vacancies; they are also marked with circles
in c.  e, f) Line profiles along the lines in a) and b), respectively. STS measured
above V$_\mathrm{O}$ and above regular 5-fold coordinated Ti$_\mathrm{5c}$
surface atoms of g) rutile and h) anatase.  Photoemission spectra (\textit{h}$\nu = 120$~eV, dashed lines) are included for comparison. Rutile at $T=78$~K, anatase at $T=6$ K. }
\end{figure}

\begin{figure}[t]
    \begin{center}
        \includegraphics[width=0.95\columnwidth]{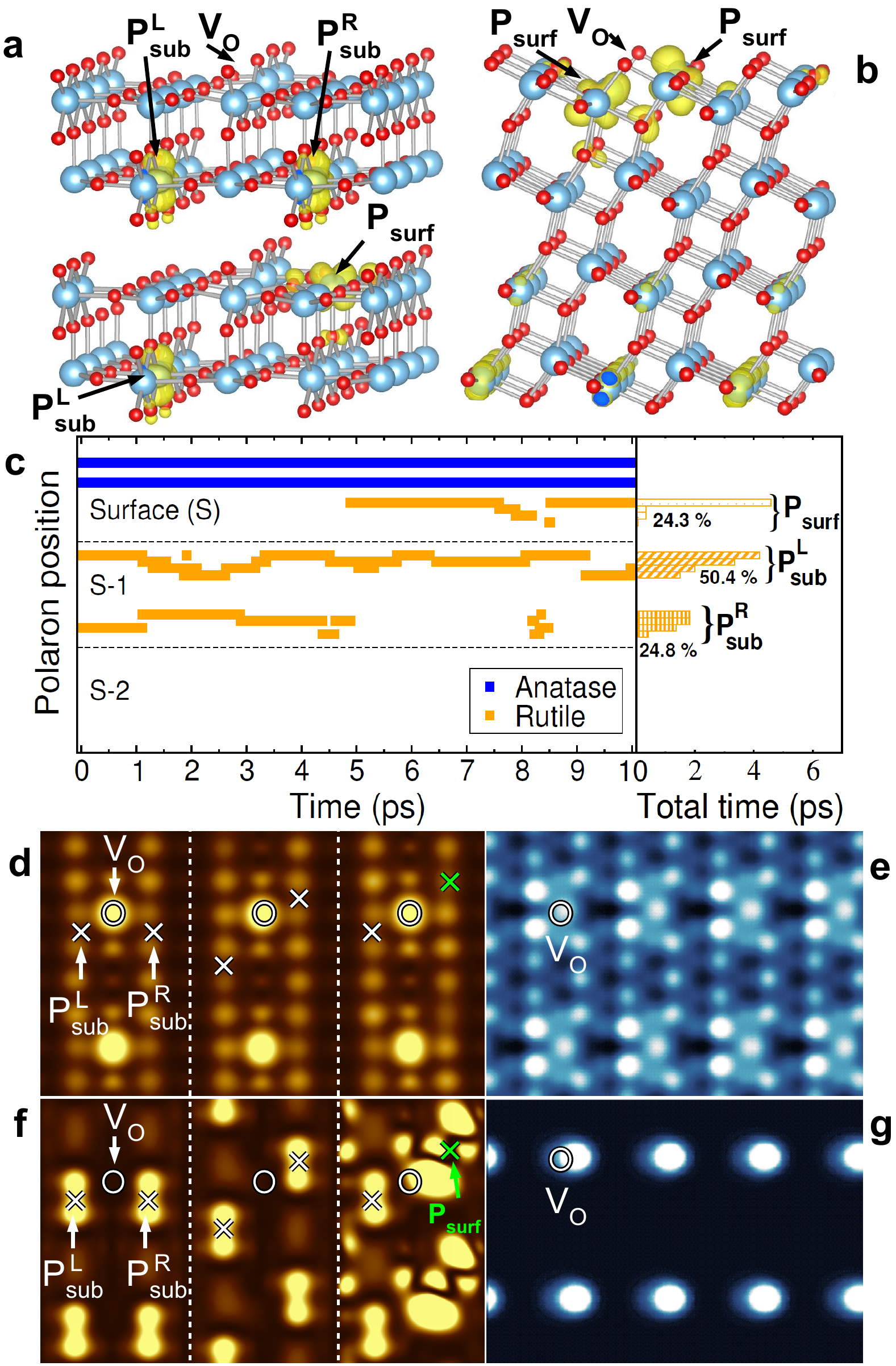}
    \end{center}
    \caption{
Surface calculations of excess electrons in rutile (110)
and anatase (101) donated by a surface V$_\mathrm{O}$.
In rutile, small polarons can assume many energetically almost equivalent positions.
a) Configurations with two subsurface polarons P$_\mathrm{sub}$ (top) and one
P$_\mathrm{sub}$ plus one surface polaron P$_\mathrm{surf}$ (bottom). b)
The only stable configuration at the anatase surface, with both excess electrons bound
to the surface V$_\mathrm{O}$.
c) Polaron dynamics in rutile (orange) and
anatase (blue) and corresponding statistical analysis for rutile.  The small polarons
stay mostly in the subsurface (S-1) below the vacancy
and at surface (S) Ti$_\mathrm{5c}$ sites. 
(d)-(g) Calculated STM images for empty (d, e) and filled (f, g) states in rutile (left, the three most frequent
polaronic configurations according to the MD analysis) and anatase (right, the electron is always trapped at the surface V$_\mathrm{O}$).}
\end{figure}

STM/STS measurements for rutile~(110) and anatase~(101) are compared in 
Fig.~2.  
As in previous work on rutile (110) \cite{20,21}, 
we take advantage of surface O vacancies (V$_\mathrm{O}$) 
that readily form under standard preparation conditions \cite{5} 
to provide excess charge (donors).  At anatase~(101), V$_\mathrm{O}$s 
migrate to the bulk at temperatures  as low as 200 K, 
but surface V$_\mathrm{O}$s  can be created non-thermally \cite{22}. Here subsurface 
V$_\mathrm{O}$s were pulled to the surface using the field 
of the STM tip as shown in ref. \cite{23}.

For rutile [Fig.~2(c)] the filled local density of states (LDOS) directly 
at the V$_\mathrm{O}$ is small; most of the current comes 
from the rows of 5-coordinated surface Ti$_\mathrm{5c}$
atoms. 
This is even more apparent when scanning at very close tip-sample distances (see 
lower part of the image in Fig.~2(c), and the Supplement for details). In contrast, 
the electrons stay at the vacancy at anatase [Fig.~2(d)], as suggested
by the vanishing LDOS at the rest of the surface.

Point tunneling spectra were measured at various distances from the 
surface V$_\mathrm{O}$s on rutile and anatase, see Figs.~2(g,h).  
The STS peak positions agree well with  photoemission spectra taken from our samples as well as other published data \cite{5,24}.  The  CBM is located just above  $E_{\mathrm{F}}$, as expected for reduced TiO$_2$;  band bending does not play a significant role. In 
STS on rutile, the polaronic band-gap-state is found $0.7 \pm 0.1$~eV 
below $E_\mathrm{F}$; again, the spectra are very similar 
when taken either directly at the vacancy or at the Ti$_\mathrm{5c}$ 
rows. On anatase, the gap state is found at $1.0\pm0.1$~eV below the 
Fermi-level \cite{24}, and it is strictly localized 
at the vacancy.  When the point spectrum is measured away from the V$_\mathrm{O}$s 
at anatase~(101), this gap state is not detected. 

Previous calculations on rutile showed that many polaronic configurations 
have almost identical total energies [Fig.~3(a)] \cite{14,25,26}. 
In our first principles Molecular Dynamics (MD) calculations, closely following Ref. \onlinecite{14}, the polarons also hop 
rapidly among lattice Ti sites [Fig.~3(c)]. Mostly ($\sim$75\% of the 
time) they stay in the first subsurface layer; occasionally ($\sim$25\%) 
they move to the surface Ti$_\mathrm{5c}$ sites \cite{14}. Calculated empty- 
and filled-states images for three such configurations are shown in Figs.~3(d,f). 
The polaron position does not affect the empty-states image, whereas it is apparent 
when imaging filled states [crosses in Fig.~3(f,g)]. 
In STM on rutile at $T= 78$~K, we measure a weighted 
average of the polaronic configurations, where both the surface and the subsurface 
small polarons contribute. Below  $T= 20$~K, we observed the complete absence
of conductivity suggesting that the polarons freeze in. 

In stark contrast to rutile, however, a surface V$_\mathrm{O}$ on 
anatase gives rise to immobile electrons pinned at Ti sites just 
at the V$_\mathrm{O}$\cite{Cheng2009} with a high $E_{\rm EL}$ 
of $\sim 1$~eV  [Figs.~3(b,c)]; these electrons are observed in filled-states 
STM (compare Figs.~2d and 3g).  In total-energy calculations any attempt to move 
the electron to a different bulk Ti lattice site resulted in an 
unstable, high-energy configuration. The electron localization close to the anatase surface
V$_\mathrm{O}$ is facilitated by high local structural 
flexibility of the lattice near the vacancy, decreasing the strain energy $E_{\rm ST}$. 

\begin{figure}
    \begin{center}
        \includegraphics[width=0.9\columnwidth,clip=true]{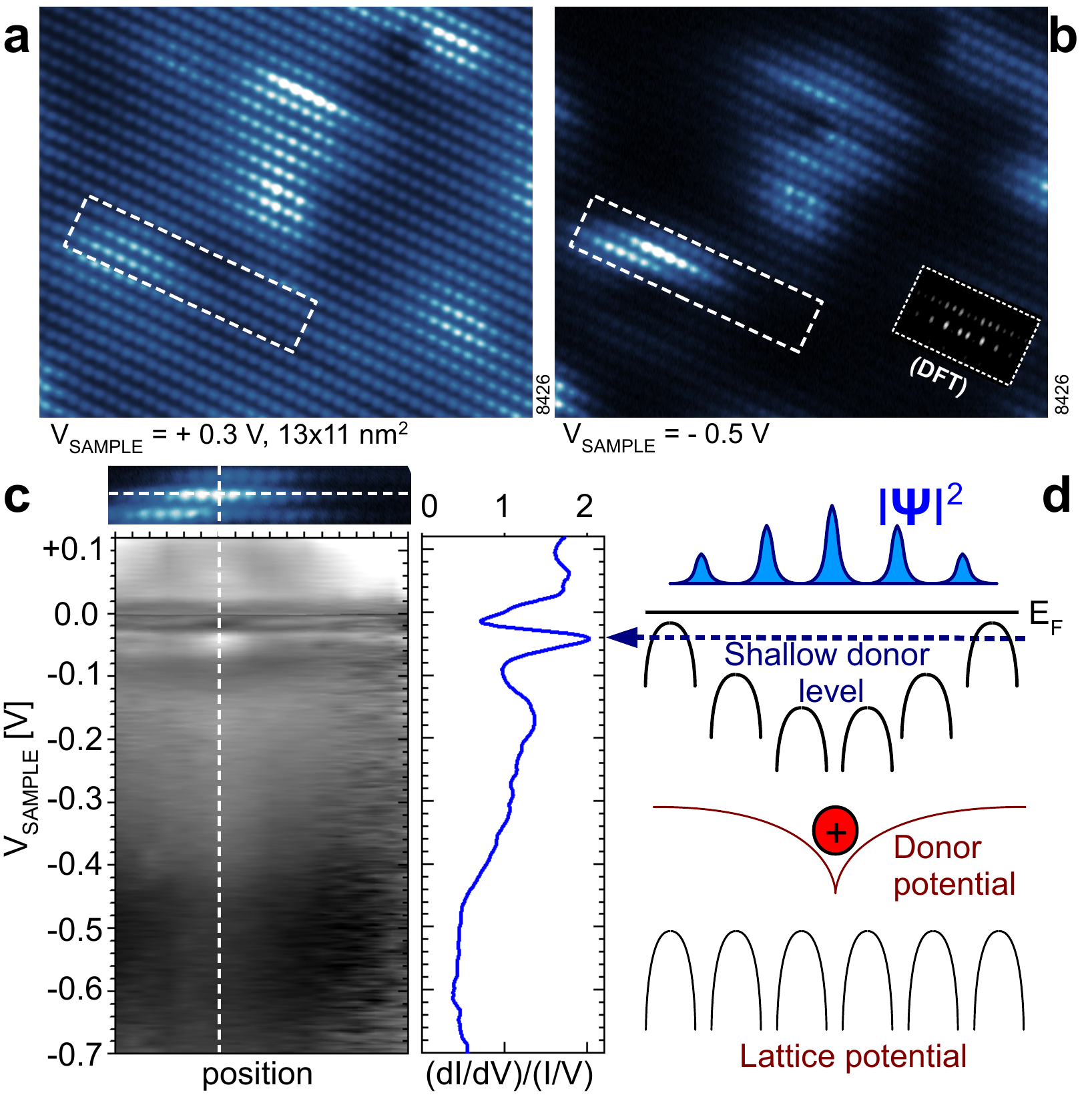}
    \end{center}
    \caption{
Shallow donor state at Nb-doped anatase~(101). a) Empty- and b) filled-states
STM images taken at $T=6$~K (constant height). The bright areas are in the vicinity
to subsurface dopants. The inset ``DFT'' in b shows a calculated STM image of the
large polaron. The large polaron wavefunction is more extended along the 
[010] (12-25~{\AA}), then along the [101] (4-8~{\AA}) direction.
c) Spatially resolved plot of (dI/dV)/(I/V) along the blue dashed
line. (Note the different energy scale as compared to Fig. 2.) d) Model of a "large
polaron", stabilized at the subsurface donor atom.
}
\end{figure}

When electrons are introduced via a dopant that modifies the lattice structure 
only slightly (See SI), polaron formation remains favorable in rutile, whereas spatially more extended
solutions are preferred in anatase. This is reflected in the 4 orders 
of magnitude larger conductivity of Nb--doped anatase compared to Nb--doped rutile: 
anatase exhibits metal--like temperature-dependence of the conductivity, whereas rutile retains a semiconductor-like character \cite{10,27}. 
Our DFT+U calculations suggest that  the higher conductivity in Nb--doped anatase is 
due to the absence of  localized small polarons, whereas in rutile -- with the same $U$ -- small polarons are formed.
The STM results are again entirely consistent with this prediction.
In an anatase sample doped with $\sim 1$\% Nb \cite{23}, fairly extended bright regions in the 
STM images with a measurable  density of states below $E_F$ 
(Fig.~4) are visible. The Nb dopants were distributed unevenly in our sample.  Fig. 4 shows a region with low concentration.
Spatially resolved STS [Fig.~4(c)] images reveal a peak at $(-40 \pm 10)$~meV and the
absence of any gap state at $-1$ eV. The bright regions shown in Figs.~4(a,b) are stable in time and do not migrate 
within a temperature range from 6 to 
78~K, suggesting that the electron is stabilized at the positively-charged  subsurface donor,
most likely Nb.
Fig.~4(d) shows a simple model. The ionized donor creates a quantum 
well, where the electron occupies a single energy level. The electron wave function 
is spread over several unit cells around the donor, and is modulated by 
the periodic potential of the crystal lattice; a calculated STM image of a slab 
with a Nb impurity [inset Fig.~4(b)] agrees well with the experiment. For details see the Supplement.
In recent ARPES measurements \cite{19} 
a similar peak at $(40 \pm 10)$~meV below E$_\mathrm{F}$ was attributed to a "large polaron". 
The distinction between a shallow, delocalized donor level and a large polaron
is subtle. DFT+U calculations reproduce the measured STM image only when lattice
relaxations (polaronic effects) are taken into account.
In STM images the density distribution has an anisotropic shape, with spatial 
extensions of $\Delta r{[010]}=12-25$~\AA~  
and $\Delta r{[101]}=4-8$~\AA. This agrees well with Fr\"ohlich's model 
for large polarons \cite{1},  from which we obtain $\Delta r{[010]}=19.0$~\AA,~ 
and $\Delta r[101]=3.5$~\AA\ related to the anisotropy of the 
screening and effective masses (see Supplement). Thus, this state is similar to a large polaron, but cannot move through the crystal like a true polaron.

Our study illustrates the basic principles of excess electron behavior in the model
oxide TiO$_2$.  The different stacking of octahedrally-coordinated Ti in the two
polymorphs, and the resulting subtle differences in the electronic structure around
the CBM, provide for a higher energy gain upon small polaron formation in rutile than
in anatase.  
Filled-states STM clearly shows the various cases:  
Small polarons that readily hop in rutile, electrons trapped at surface V$_\mathrm{O}$s in anatase, and 
spatially extended shallow donor states in Nb-doped anatase.
 STS also shows distinct signatures, with an apparent deep ($\sim$eV)
polaronic state for rutile, yet a much shallower state (40 meV) at dopants of anatase.
The latter shows spectroscopic similarities with recent
ARPES results \cite{19} and its lateral extension is
well-described by Fr\"ohlich's  theory for large polarons.
These experimental results prove electron de/localization for anatase/rutile \textit{surfaces.} Recent EPR data \cite{3} indicate the same behavior in the  \textit{bulk} of these materials, in agreement with the calculations in Fig. 1(b).

Polarons are central to the often exotic behavior of oxides \cite{28} as well as
their technological applications. In the specific case of TiO$_2$,  anatase is used as an electrode in
photoelectrochemical  solar cells.  Band-like charge transport and the lack of small
polaron formation is the key requirement for increasing the cell efficiency. On the
other hand, the formation of small polarons in rutile is an
asset in catalysis, as the polaron formation is more favorable at surfaces than in the bulk, facilitating an efficient charge transfer to catalyzed species \cite{6}. In mixtures of the two TiO$_2$ phases,
anatase provides a good electron conductor that transports charge carriers to the
interface with rutile, where they are trapped \cite{30}.


{This work was supported by the ERC Advanced Research Grant `OxideSurfaces',
and by the Austrian Science Fund (FWF). Computing time provided by the Vienna Scientific Cluster is gratefully
aknowledged.}

\end{document}